\documentclass{article}
\usepackage{graphicx}
\title{%
Quasi-Particle States with Topological Quantum Numbers 
in the Mixed State of $d$-wave Superconductors
}

\author{
Tomio \textsc{KOYAMA}$
$\thanks{tkoyama@imr.tohoku.ac.jp}
 \\
Institute for Materials Research, Tohoku University, Katahira 2-1-1, \\
Aoba-ku, Sendai 980-8577
}

\begin{document}
\maketitle
\begin{abstract}
We investigate the extended quasi-particle states in the mixed state of 
$d$-wave superconductors on the basis of the Bogoliubov-de Gennes equation. 
We prove that the quasi-particle eigen-states can be classified in terms 
of new topological quantum numbers which are related to the topological 
nature of the non-trivial phases of the quasi-particles. Numerical results 
for the quasi-particle eigen-states reveal the crossover behavior from 
gapless to gapped states as the flux density $B$ increases. In the strong 
field region quantum oscillations appear in the excitation energy of the 
quasi-particles. 
\end{abstract}


\section{Introduction}
The quasi-particles in the mixed state of $d$-wave superconductors has 
attracted much attention in connection with the peculiar low-energy 
excitations under a magnetic field in high-$T_c$ cuprates~\cite{rf:1,rf:2,
rf:3,rf:4,rf:5,rf:6,rf:7,rf:8,rf:9,rf:10,rf:11}. 
In this paper we investigate the extended quasi-particle states in 
extremely type-II $d$-wave superconductors on the basis of the 
Bogoliubov-de Gennes (BdG) equation, focusing on the topological nature 
of the quasi-particles. It is well-known that the order parameter has a 
non-trivial phase in the presence of vortices, that is, the phase 
$\phi({\bf r})$ is a multi-valued function satisfying the relation, 
$\nabla\times\nabla\phi({\bf r})=2\pi\sum_i^N\delta({\bf r-R}_i){\bf e}_z$, 
in the mixed state where vortices ($\parallel {\bf e}_z$) are located at 
${\bf R}_1, {\bf R}_2,\cdots, {\bf R}_N$. Then, the BdG equation contains 
a multi-valued function, which makes the equation intractable. 
Anderson noticed that the phase $\phi({\bf r})$ in the BdG equation can 
be eliminated in terms of a singular phase transformation, by which the 
BdG equation is transformed into the one containing 
$\nabla\phi({\bf r})$~\cite{rf:7}. This is an important observation, 
because $\nabla\phi({\bf r})$ is a single-valued function, that is, the 
transformed BdG equation includes only single-valued functions. However, 
it turns out, as shown in this paper, that the topological character of 
the system is not fully incorporated into the wave functions by this 
transformation. One should note that a superconductor containing vortices 
may be considered as a multi-connected system. Then, the quasi-particles 
moving around vortices show the AB effect~\cite{rf:11}. From this fact one 
understands that the wave functions of the quasi-particles should be 
\textit{path-dependent single-valued functions} in the presence of vortices. 

In this paper we clarify the topological character of the non-trivial 
phases of the wave functions satisfying the BdG equation and show that the 
quasi-particle eigen-states have a \lq\lq hidden" quantum number which is 
related to the homotopy class of the classical orbits of the quasi-particles. 
We also develop a theoretical scheme for solving approximately the BdG 
equation in the field region $H_{c1}\ll H \ll H_{c2}$, correcting the 
incomplete treatments for the topological singularity of $\phi({\bf r})$ in 
previous works. Numerical solutions for the extended quasi-particle states 
in $d$-wave superconductors are briefly presented. It is shown that the 
gapless quasi-particle state changes to a gapped one with increasing the 
flux density. In the strong field region quantum oscillations appear 
in the excitation energy. 


\section{Formulation}

Consider the BdG equation for a $d_{xy}$-wave superconductor, 
\begin{equation}
\left[
  \begin{array}{cc}
      h_0(\textbf{r},\nabla)
                    & \hat\Delta(\textbf{r},\nabla) \\
      \hat\Delta^\ast(\textbf{r},\nabla) & -h_0^\ast(\textbf{r},\nabla)
  \end{array}
\right]
\left[
  \begin{array}{c}
     u_\alpha(\textbf{r}) \\ v_\alpha(\textbf{r})
  \end{array}
\right]
= E\left[
  \begin{array}{c}
     u_\alpha(\textbf{r}) \\ v_\alpha(\textbf{r})
  \end{array}
\right],
\end{equation}
where 
\begin{equation}
 h_0(\textbf{r},\nabla)=\frac{1}{2m}\bigl({\bf p}-\frac{e}{c}{\bf A(r)}
\bigl)^2-\varepsilon_F, 
\end{equation}
and $\hat\Delta(\textbf{r},\nabla)$ is the gap operator defined as 
\begin{equation}
\hat\Delta(\textbf{r},\nabla)=-\frac{1}{k_F^2}\bigl\{\partial_x,
\bigl\{\partial_y,\Delta(\textbf{r})\bigr \}\bigr \}
  -\frac{\textrm{i}}{4}\Delta({\bf r})\bigl\{\partial_x,\partial_y\}
\phi(\textbf{r}), 
\end{equation}
with $\{A,B\}\equiv \frac{1}{2}(AB+BA)$~\cite{rf:3}.
In eq.(3) $\phi(\textbf{r})$ is the phase of the gap function 
$\Delta({\bf r})$ and the second term is added for recovering the 
U(1)-symmetry~\cite{rf:12,rf:13}. In this paper we investigate the extended 
quasi-particle states outside the vortex cores in the field region 
$H_{c1}\ll H \ll H_{c2}$, so that we may approximate the gap function as 
\begin{equation}
 \Delta(\textbf{r})\simeq \Delta_0\textrm{e}^{\textrm{i}\phi(\textbf{r})},
\end{equation}
neglecting the spatial dependence of the gap amplitude. In the case where 
$N$ vortices ($\parallel z$-axis) are located at ${\bf R_1, \cdots, 
R_N}$ in the $xy$ plane, the phase $\phi$ in eq.(3) is dependent on the 
vortex positions, i.e., 
$\phi(\textbf{r})=\phi(\textbf{r}; {\bf R_1,\cdots, R_N})$,
and is a multi-valued function satisfying the relation, 
\begin{equation}
 \nabla\times\nabla\phi(\textbf{r}; {\bf R_1, \cdots, R_N})=2\pi\sum_{i=1}^N
\delta({\bf r-R}_i){\bf e}_z.
\end{equation}
To express the BdG equation in terms of only single-valued functions, 
removing the multi-valued function $\phi({\bf r})$, Anderson proposed a 
transformation for the wave functions 
$u_\alpha$ and $v_\alpha$ in eq.(1) as follows~\cite{rf:7}, 
\begin{equation}
  u_\alpha \rightarrow u_\alpha,\ 
  v_\alpha \rightarrow \textrm{e}^{-\textrm{i}\phi}v_\alpha, \ \ \ \ 
{\rm or}\ \ \  \ 
     u_\alpha \rightarrow \textrm{e}^{\textrm{i}\phi}
     v_\alpha \ v_\alpha \rightarrow v_\alpha.
\end{equation}
This transformation has been utilized by several authors for studying the 
low energy excitations in the mixed state of $d$-wave superconductors~\cite{ 
rf:9,rf:10}. On the other hand, Franz and Te\v{s}anovi\'{c} pointed out 
that the phase can be eliminated also by the transformation, 
$u_\alpha \rightarrow \textrm{e}^{\textrm{i}\phi_e}
u_\alpha, v_\alpha \rightarrow \textrm{e}^{-\textrm{i}\phi_h}
v_\alpha$, if $\phi_e+\phi_h=\phi$~\cite{rf:8,rf:14}. The phase factor 
$\textrm{e}^{\textrm{i}\phi}$ in eq.(1) can really be excluded from the 
BdG equation by these transformations, but we claim that these manipulations 
should not be considered as the \textit{transformations}. Let us suppose 
that the wave functions in the $N$-vortex state, 
$u_\alpha(\textbf{r}),v_\alpha(\textbf{r})$, are expressed as 
\begin{equation}
\left\{
  \begin{array}{l}
u_\alpha(\textbf{r})=\tilde u_{n\mu}(\textbf{r})\textrm{e}^{\textrm{i}
(\frac{1}{2}-\mu)\phi({\bf r;R_1,\cdots, R_N})} \\
v_\alpha(\textbf{r})=\tilde v_{n\mu}(\textbf{r})\textrm{e}^{-\textrm{i}
(\frac{1}{2}+\mu)\phi({\bf r;R_1,\cdots, R_N})} 
  \end{array}, 
\right.
\end{equation}
when the non-trivial phases in the wave functions are explicitly extracted. 
In eq.(7) $\mu$ is assumed to be a \textit{half and integer}, namely, 
$\mu=\pm \frac{1}{2},\pm \frac{3}{2}, \cdots$. Note that the phases in 
eq.(7) satisfy the Franz-Te\v{s}anovi\'{c} condition mentioned 
above ($\phi_e+\phi_h=\phi$) and eq.(7) is also reduced to the Anderson's 
transformation (6) for $\mu=\pm\frac{1}{2}$ if eq.(7) is considered as 
the transformation, $(u_\alpha, v_\alpha) \rightarrow (\tilde u_\alpha, 
\tilde v_\alpha)$. Hence, it is clear that the phase factor ${\rm e}^{
{\rm i}\phi}$ is eliminated in the equation for $\tilde u_\alpha, \tilde 
v_\alpha$. The new equation for $\tilde u_\alpha$ and $\tilde v_\alpha$ 
contains explicitly a number $\mu$, so that the energy eigenvalues 
explicitly depend on $\mu$. Furthermore, it turns out from the calculations 
shown later that the energy eigen-states can be classified in terms of $\mu$. 
From these results one can interpret $\mu$ as a \textit{quantum number 
specifying the quasi-particle eigen-states} in the $N$-vortex state. Then, 
a set of quantum numbers $\alpha$ in eq.(7) is expressed as $\alpha=(n,\mu)$, 
where $n$ denotes symbolically the other quantum numbers. 

To be convinced of our statement, let us now study the topological character 
of the wave functions. We introduce the operator $\mathcal{P}(C)$ which 
moves a function along a path $C$~\cite{rf:15}.
Let us take a closed path $C_i$ around the $i$th vortex at ${\bf R}_i$ for 
$C$. Suppose that $C_i$ does not include any other vortices inside it 
(see Fig.1). 
Then, the operation of $\mathcal{P}(C_i)$ on the phase 
$\phi({\bf r})$ induces the shift, 
\begin{equation}
\mathcal{P}(C_i)\phi({\bf r:R_1,\cdots,R_N})
=\phi({\bf r:R_1,\cdots,R_N})+2\pi.
\end{equation}
%
%
%
Suppose that the wave functions, $u_\alpha({\bf r})$ and 
$v_\alpha({\bf r})$, have non-trivial phases, $\Theta_u({\bf r})$ and 
$\Theta_v({\bf r})$, i.e., $u_\alpha({\bf r})=\tilde u_\alpha({\bf r})
\textrm{e}^{\textrm{i}\Theta_\alpha^u({\bf r})}$ and $v_\alpha({\bf r})
=\tilde v_\alpha({\bf r})\textrm{e}^{\textrm{i}\Theta_\alpha^v({\bf r})}$. 
Since $u_\alpha({\bf r})$ and $v_\alpha({\bf r})$ should be \textit{
single-valued functions}, we have in general
\begin{equation}
\mathcal{P}(C_i)u_\alpha=u_\alpha\textrm{e}^{\textrm{i}
2\pi\ell}, \ \ 
\mathcal{P}(C_i)v_\alpha=v_\alpha\textrm{e}^{\textrm{i}2\pi m},
\end{equation}
with $\ell$ and $m$ being integers. Then, eq.(5) indicates the relations, 
\begin{equation}
\mathcal{P}(C_i)\Theta_\alpha^u=\Theta_\alpha^u + 2\pi\ell, 
\ \ \ \ 
\mathcal{P}(C_i)\Theta_\alpha^v=\Theta_\alpha^v + 2\pi m,
\end{equation}
under the conditions, $\mathcal{P}(C_i)\tilde u_\alpha=\tilde u_\alpha$ and 
$\mathcal{P}(C_i)\tilde v_\alpha=\tilde v_\alpha$. Furthermore, from the 
gap equation, 
$\Delta({\bf r})\sim\sum_\alpha f_\alpha\tilde u_\alpha\tilde v_\alpha^\ast
\textrm{e}^{\textrm{i}(\Theta_\alpha^u-\Theta_\alpha^v)}$ (see Appendix A), 
one may assume the equality, 
\begin{equation}
 \phi({\bf r})=\Theta_\alpha^u({\bf r}) - \Theta_\alpha^v({\bf r}),
\end{equation}
Then, from the operation of $\mathcal{P}(C_i)$ on both sides of eq.(11) it 
follows,
\begin{equation}
 \mathcal{P}(C_i)\phi=\mathcal{P}(C_i)
 \bigl(\Theta_\alpha^u - \Theta_\alpha^v\bigr)
 =\phi + 2\pi(\ell-m). 
\end{equation}
Comparing this relation with eq.(8), one finds 
\begin{equation}
\ell-m=1, \ \ \ \textrm{or} \ \ \ 
   \ell=\frac{1}{2}-\mu, \ \  m=-(\frac{1}{2}+\mu), 
\end{equation}
with $\mu$ being a half and integer. 
Since these relations are simultaneously satisfied 
when 
\begin{equation}
\left\{
 \begin{array}{ll}
   \Theta_\alpha^u({\bf r})= & (\frac{1}{2}-\mu)\phi({\bf r}) \\
   \Theta_\alpha^v({\bf r})= &-(\frac{1}{2}+\mu)\phi({\bf r}) 
 \end{array}
\right.,
\end{equation}
we obtain eq.(7). We can also show that the gap equation expressed in terms 
of the wave functions (7) yields $\phi({\bf r})$ for the phase of the order 
parameter (see Appendix A). Note that eq.(7) holds for $N=1$, i.e., 
the single vortex state. As is well-known, the wave functions in the 2D 
single vortex state are expressed as 
\begin{equation}
\left\{
  \begin{array}{ll}
       u_{n\mu}(r,\theta)=\tilde u_{n\mu}(r)
           \textrm{e}^{\textrm{i}(\frac{1}{2}-\mu)\theta} \\
       v_{n\mu}(r,\theta)=\tilde v_{n\mu}(r)
           \textrm{e}^{-\textrm{i}(\frac{1}{2}+\mu)\theta}
  \end{array}
\right.,
\end{equation}
where $(r,\theta)$ is the polar coordinates~\cite{rf:16}, which indicates 
that the phase $\phi({\bf r})$ is expressed in terms of the geometrical 
angle $\theta$, i.e., $\phi({\bf r})=\theta$, in this state and then $\mu$ 
($=\pm \frac{1}{2}, \frac{3}{2}, \cdots)$ accords with the quantum number 
for the 2D angular momentum, though in general $\mu$ is a topological one. 
Thus, our conjecture (7) is consistent with the well-known result in the 
single vortex state. Note that $\mu$ is still a rigorous quantum number 
in the system with a curved vortex, in which the angular momentum is not 
conserved.


Let us now study the BdG equation (1), using eq.(7). In the present 
$d$-wave case we must be careful about the order of differentiations in 
eq.(3), because $\partial_x\partial_y\phi({\bf r})\not= 
\partial_y\partial_x\phi({\bf r})$ in the vortex state~\cite{rf:17}. 
To avoid the ambiguity in the order of differentiations we symmetrize the 
differential operations in eq.(3) as 
$\partial_x\partial_y\textrm{e}^{\textit{i}\phi({\bf r})}\rightarrow 
\{\partial_x,\partial_y\}\textrm{e}^{\textit{i}\phi({\bf r})}$. Then, 
substitution of eq.(7) into eq.(1) yields  
\begin{equation}
\left[
  \begin{array}{cc}
   \hat h_0({\bf r},\nabla;\mu) & \Pi({\bf r},\nabla;\mu) \\
\Pi^\ast({\bf r},\nabla;-\mu) & 
   -\hat h_0^\ast({\bf r},\nabla;-\mu)
  \end{array}
\right]
\left[
  \begin{array}{c}
    \tilde u_{n\mu} \\
    \tilde v_{n\mu}
  \end{array}
\right]
= E\left[
  \begin{array}{c}
    \tilde u_{n\mu} \\
    \tilde v_{n\mu}
  \end{array}
\right],
\end{equation}
where 
\begin{equation}
  \hat h_0({\bf r},\nabla;\mu)=-\frac{\hbar^2}{2m}\bigl[\nabla+\textrm{i}
\frac{m}{\hbar}{\bf v}_s-\textrm{i}\mu\nabla\phi\bigr]^2 - \varepsilon_F,
\end{equation}
${\bf v}_s$ is the superfluid velocity defined by 
\begin{equation}
{\bf v}_s=\frac{\hbar}{m}\bigl(\nabla\phi-\frac{2e}{\hbar c}{\bf A}\bigr),
\end{equation}
and the off-diagonal component, $\Pi({\bf r},\nabla;\mu)$, is given as 
\begin{equation}
\Pi({\bf r},\nabla;\mu)
=-\frac{\Delta_0}{k_F^2}\Bigl(
      -\mu^2\partial_x\phi\cdot\partial_y\phi
     -\textrm{i}\mu\{\partial_x,\partial_y\}\phi
  -\textrm{i}\mu\bigl(\partial_y\phi\cdot\partial_x
             +\partial_x\phi\cdot\partial_y\bigr)
  +\partial_x\partial_y\Bigr).
\end{equation}
To perform explicit calculations we introduce an approximation in the 
following. We investigate the region $H_{c1}\ll H \ll H_{c2}$ in the 
extremely-type II superconductors of $\kappa\gg 1$, i.e., $\xi\ll\lambda_L$. 
In this field region one can assume the relation, $\xi\ll d \ll \lambda_L$, 
where $d$ is the lattice constant of the flux-line-lattice. The 
spatial-dependent flux density in this region, ${\bf B}({\bf r})$, can be 
split into the average value, ${\bf B}_{\rm av}$, and the deviation from it, 
$\delta{\bf B}({\bf r})$, i.e., ${\bf B}({\bf r})={\bf B}_{\rm av}+\delta
{\bf B}({\bf r})$. Since the lattice constant $d$ is much shorter than the 
London penetration depth $\lambda_L$ in this field region, the vortex currents 
flowing around vortices are heavily overlapped and, as a result, canceled out, 
which indicates the relation, $|\delta{\bf B}({\bf r})|\ll 
|{\bf B}_{\rm av}|$. Consequently, the spatial-dependent term $\delta{\bf B}
({\bf r})$ is neglected in the 0th order approximation, ${\bf B}({\bf r})
\sim {\bf B}_{\rm av}$. In this paper we confine ourselves to the lowest 
order approximation. Then, in this approximation we have the relation from 
the Maxwell equation as follows, 
\begin{equation}
 \nabla\times{\bf B}({\bf r})=\frac{4\pi e^\ast}{c}\rho_s({\bf r})
{\bf v}_s({\bf r})\simeq \nabla\times{\bf B}_{\rm av}=0, 
\end{equation}
where $\rho_s({\bf r})$ is the local superfluid density. Since 
$\rho_s({\bf r})\simeq$ const.$\not=0$ outside the vortex cores, one can 
assume ${\bf v}_s\simeq 0$ in the present approximation. Then, noting 
eq.(18), we find the approximate relation, 
\begin{equation}
\nabla\phi({\bf r})\simeq \frac{2e}{\hbar c}{\bf A}({\bf r})
\simeq \frac{2e}{\hbar c}{\bf A}_0({\bf r}), 
\end{equation}
where the vector potential ${\bf A}_0({\bf r})$ is related to the average 
flux density as ${\bf B}_{\rm av}=\nabla\times {\bf A}_0({\bf r})$. 
Since ${\bf A}_0({\bf r})=(0, Bx,0)$ with $B=|{\bf B}_{\rm av}|$ 
in the Landau gauge ($B<H$), we arrive at the explicit expression for 
the non-trivial phase as 
\begin{equation}
     \partial_x\phi= 0,  \ \ \ \ 
     \partial_y\phi= \frac{2e}{\hbar c}Bx=\frac{2\pi}{\phi_0}Bx,
\end{equation} 
where $\phi_0$ is the unit flux, $\phi_0=\frac{hc}{e^\ast}$. 
Eq.(22) indicates the relation, 
\begin{equation}
  \partial_x\partial_y\phi=\frac{2e}{\hbar c}B, \ \ \ \ 
  \partial_y\partial_x\phi=0,
\end{equation}
that is, the differential operations for $\phi$ do not commute, 
$\partial_x\partial_y\phi\not=\partial_y\partial_x\phi$, as expected. 
Then, we have the topological singularity in the present approximation, 
\begin{equation}
(\nabla\times\nabla\phi({\bf r})|_z
=(\partial_x\partial_y-\partial_y\partial_x)\phi({\bf r})=2\pi m
\end{equation}
noting $B=m\phi_0$ with $m$ being an integer. This result corresponds to 
that in the \lq\lq continuity-approximation" for the singularity, i.e., 
$(\nabla\times\nabla\phi({\bf r}))_z \simeq 2\pi\langle \sum_i^N \delta
({\bf r-R_i})\rangle_{\rm av.}=2\pi m$~\cite{rf:18}.


Let us now solve the BdG equation, using the above approximation. 
Under the approximation eq.(16) can be rewritten in the 2D case as follows, 
\begin{equation}
\left[
  \begin{array}{cc}
   \hat h_0(P,Q) & \Pi(P,Q;\mu) \\
%
    \Pi^\ast(P,Q;-\mu)&  -\hat h_0(P,Q)
   \end{array}
\right]
\left[
  \begin{array}{c}
     \tilde u_{n\mu} \\ \tilde v_{n\mu}
   \end{array}
\right]
=E\left[
  \begin{array}{c}
     \tilde u_{n\mu} \\ \tilde v_{n\mu}
   \end{array}
\right], 
\end{equation}
where 
\begin{equation}
 \hat h_0(P,Q)=\mu\omega_c\bigl(Q^2+P^2\bigr)-\varepsilon_F,
\end{equation}
and 
\begin{equation}
 \Pi(P,Q;\mu)=\frac{\Delta_0}{2\varepsilon_F}\hbar\omega_c\bigl[
    \textrm{i}\mu + 2\frac{\mu}{\hbar}PQ \bigr]. 
\end{equation}
Here, $\omega_c=\frac{eB}{mc}$ is the cyclotron frequency and 
\begin{equation}
 Q=\bigl(\mu\frac{2e}{c}B\bigr)^{-\frac{1}{2}}p_x, \ \ \ \ 
  P=\bigl(\mu\frac{2e}{c}B\bigr)^{-\frac{1}{2}}
   \bigl(p_y-\mu\frac{2e}{c}Bx\bigr), 
\end{equation} 
with $p_\alpha=-\textrm{i}\hbar\partial_\alpha$. Note that $Q$ and $P$ 
satisfy the commutation relation, $[Q,P]=\textrm{i}\hbar$. Thus, the BdG 
equation is expressed in terms of only Hermitian operators and is 
gauge-invariant. we remark that symmetrizing the differential operations 
in eq.(3) is essential for obtaining the above physically-acceptable results. 
Now let us introduce the raising and lowering operators, $a^\dagger$ and $a$, 
as 
  $Q=\textrm{i}\sqrt{\frac{\hbar}{2}}(a-a^\dagger)$ and 
  $P=\sqrt{\frac{\hbar}{2}}(a+a^\dagger)$, 
noting the above commutation relation for $Q$ and $P$. Then, eq.(25) is 
rewritten as 
\begin{equation}
\left[
  \begin{array}{cc}
   2\mu\bigl(a^\dagger a + \frac{1}{2}\bigr)-\varepsilon_F^\prime & 
   -\frac{\Delta_0}{2\varepsilon_F}\mu(a^2-a^{\dagger 2}) \\
   \frac{\Delta_0}{2\varepsilon_F}\mu(a^2-a^{\dagger 2}) & 
   -2\mu\bigl(a^\dagger a + \frac{1}{2})+\varepsilon_F^\prime 
   \end{array}
\right]
\left[
  \begin{array}{c}
     \tilde u_{n\mu} \\ \tilde v_{n\mu}^\prime
   \end{array}
\right]
=\frac{E_{n\mu}}{\hbar\omega_c}
\left[
  \begin{array}{c}
     \tilde u_{n\mu} \\ \tilde v_{n\mu}^\prime
   \end{array}
\right], 
\end{equation}
where $\tilde v_{n\mu}^\prime=-\textrm{i}\tilde v_{n\mu}$ and 
$\varepsilon_F^\prime=\frac{\varepsilon_F}{\hbar\omega_c}$. 
To solve eq.(29) numerically it is convenient to introduce the expansion in 
terms of the Landau levels, 
\begin{equation}
  \tilde u_{n\mu}=\sum_{k=0}^\infty \alpha_{2k+1}^{n\mu}|k\rangle, \ \ \ \ 
  \tilde v_{n\mu}^\prime=\sum_{k=0}^\infty \alpha_{2k+2}^{n\mu}|k\rangle, 
\end{equation}
where $|k\rangle$ is the $k$th Landau level, i.e.,
$a|k\rangle=\sqrt{k}|k-1\rangle$ and 
$a^\dagger|k\rangle=\sqrt{k+1}|k+1\rangle$. It is seen that the off-diagonal 
components in eq.(29) induce the mixing between the Landau levels, that is, 
the Landau levels are not the eigen-states in the present $d$-wave case. 
Eq.(29) also indicates that the quasi-particle eigen-states in the mixed 
state of 2D $d$-wave superconductors can be classified in terms 
of two quantum numbers, namely $\mu$, which has the topological origin, 
and $n$, which is reduced to the Landau-level index in the normal state 
($\Delta_0\rightarrow 0$). The quantum number $\mu$ is a rigorous quantum 
number in the mixed state of type II superconductors, which is independent 
of approximations used in this paper. 
From the above results we finally obtain the coupled equations for the 
coefficients, $\alpha_k^{n\mu}$, as
\begin{equation}
\frac{1}{4}\mu\tilde\Delta_0\sqrt{(k-3)(k-1)}\alpha_{k-3}^{n\mu}
+[\mu k-\varepsilon_F^\prime]\alpha_k^{n\mu}
%
-\frac{1}{4}\mu\tilde\Delta_0\sqrt{(k+1)(k+3)}
\alpha_{k+5}^{n\mu}=\frac{E}{\hbar\omega_c}\alpha_k^{n\mu},
\end{equation}
for $k=1,3,5,\cdots$, and 
\begin{equation}
-\frac{1}{4}\mu\tilde\Delta_0\sqrt{(k-4)(k-2)}\alpha_{k-5}^{n\mu}
-[\mu (k-1)-\varepsilon_F^\prime]\alpha_k^{n\mu}
+\frac{1}{4}\mu\tilde\Delta_0\sqrt{k(k+2)}
\alpha_{k+3}^{n\mu}=\frac{E}{\hbar\omega_c}\alpha_k^{n\mu},
\end{equation}
for $k=2,4,6,\cdots$, with $\tilde\Delta_0=\frac{\Delta_0}{\varepsilon_F}$.

%
%
%

\section{Numerical results and discussions}

Let us now present numerical solutions of eqs.(31) and (32) obtained 
by a numerical diagonalization method. Fig.2 shows the positive energy 
eigen-values for the quantum numbers, $\mu=\frac{1}{2},\frac{3}{2},
\frac{5}{2}$. The parameter values in these calculations are chosen as 
$\varepsilon_F=200$ meV, $\Delta_0=10$ meV, and the free electron mass 
$m_e$ is assumed in the cyclotron frequency, i.e., $\omega_c=\frac{eB}{m_ec}$. 
As seen in this figure, we have two quasi-particle eigen-states with 
nearly-zero eigen-values. 
These states appears in the case of $\hbar\omega_c\ll \Delta_0$ in our 
calculations. The existence of the zero-energy state has been predicted in 
the approximate calculations in ref.[7]. These zero-energy states are 
formed mainly by the quasi-particles lying in the nodal directions 
in the $k$-space. It is also seen that the two or four consecutive levels 
have close eigen-values in the low energy region. This result comes from 
that the state with index $k$ does not couple with that with $k+\ell$ of 
$\ell\leq 2$ or $\ell\leq 4$, as seen in eqs.(31) and (32). 
In Fig.3 we plot the energy eigen-values in the $\mu=\frac{1}{2}$ band 
for three values of $B$. Note that the relation $\hbar\omega_c\ll \Delta_0$ 
holds in these cases, since $\hbar\omega_c\sim 1.2$meV$\ll\Delta_0$ for 
$B=10$ T in the 
present choice of the parameter values. Then, the energy levels are 
essentially the same in this field region, except that the dispersion 
becomes steeper as $B$ increases. 

%
%
%

To investigate the crossover behavior due to the variation of the flux 
density $B$ from $\hbar\omega_c\ll \Delta_0$ to $\hbar\omega_c \gg \Delta_0$ 
we calculate the lowest excitation energy as a function of $B$, choosing a 
small value for $\Delta_0$. Fig.4 shows the $B$-dependence of $E_{n\mu}$ 
with $n=1,\mu=\frac{1}{2}$ in the case of $\Delta_0=2$ meV. In this choice 
of the parameter values we find $\hbar\omega_c=\Delta_0$ for $B\simeq 16$T.
As seen in this figure, the lowest energy eigen-state has zero-energy in 
the weak field region of $\hbar\omega_c\ll \Delta_0$, but it abruptly 
changes to a gapped state above some value of $B$, which indicates that 
the quasi-particle eigen-states acquire a full energy-gap in the strong 
field region, though the symmetry of the gap function is the same as in 
the Meissner state. This result may be intuitively understood in the 
following way. The quasi-particles lying in the nodal directions with 
$p_x\sim p_F$ cannot turn by the effect of a weak Lorentz force when the 
pair-potential barrier is high. Then, in the weak filed region the 
quasi-particle states in the nodal directions are expected to be gapless. 
However, in the strong field region where the Lorentz force is strong 
enough, i.e., $\hbar\omega_c\sim \Delta_0$. the quasi-particles in the 
nodal directions can run through the pair-potential barrier. In order to 
excite the quasi-particles going through the barrier of the pair-potential 
we need a finite excitation energy, which leads to a finite energy gap. 
It has been proved in the GL limit, in which the Landau-level splitting 
is completely neglected, that the $d_{xy}$-component is induced in the 
mixed state of $d_{x^2-y^2}$-wave superconductors~\cite{rf:19}, which 
also creates fully-gapped quasi-particle states. However, the origin of 
the energy gap in the GL limit is completely different from that in the 
present theory. This result seems consistent with the experimental 
observation for the thermal conductivity in the mixed state~\cite{rf:6}. 
In the strong field region of $\hbar\omega_c > \Delta_0$, namely $B>16$T 
in the present case, the lowest excitation energy shows oscillatory 
behavior in the $B$-dependence, as seen in Fig.4. These oscillations reflect 
the field dependence of the diagonal components of eq.(29), that is, they 
are the quantum oscillations which lead to the de Haas-van Alphen effect. 
Thus, our theory for the extended quasi-particles in the mixed state of 
$d$-wave superconductors can describes the crossover behavior from the 
gapless phase in the weak field region to the gapped phase showing the 
quantum oscillations in the strong field region.


In summary we investigated the extended quasi-particle states of $d$-wave 
superconductors in the field region, $H_{c1}\ll H \ll H_{c2}$, on the basis 
of the BdG equation. We found new topological quantum numbers which classify 
the quasi-particle eigen-states in the mixed state of type-II superconductors. 
The phase of a qausi-particle should be path-dependent in the presence of 
vortices, which is essentially the same as in a multi-connected system 
showing the Aharnnov-Bohm effect. Hence, one understands that the quantum 
numbers carrying information about the homotopy class of the orbit of 
the quasi-particle appear in the mixed state of type II superconductors. 
We also presented numerical solutions of the BdG equation in the field 
region, $H_{c1}\ll H \ll H_{c2}$. It was shown that the quasi-particle 
states show the crossover behavior from gappless to gapped states as the 
flux density increases. In the strong field region of $\hbar\omega_c>
\Delta_0$ the quantum oscillations appear in the excitation energy.

\section*{Acknowledgements}
The author would like to thank Dr. M. Machida for useful discussions. 


\appendix
\section{Appendix }

In this Appendix we derive the gap equation in the present 
$d_{xy}$-superconductors. The gap function in unconventional 
superconductors has generally the form, 
$$
\Delta({\bf r,r}^\prime)=V({\bf r-r}^\prime)\langle
\psi_\uparrow({\bf r})\psi_\downarrow({\bf r}^\prime)\rangle
$$
\begin{equation}
=V({\bf r-r}^\prime)\sum_i\Bigl\{
u_i({\bf r})v_i^\ast({\bf r}^\prime)\bigl[1-f_{\rm F}(E_i)\bigr]
     -v_i^\ast({\bf r})u_i({\bf r}^\prime)f_{\rm F}(E_i)\Bigr\}, 
\end{equation}
where $V({\bf r-r}^\prime)$ denotes the interaction between superconducting 
electrons. We introduce the Fourier transformation for the relative 
coordinates, ${\bf s}={\bf r}-{\bf r}^\prime$ as 
\begin{equation}
\Delta({\bf r,r}^\prime)=\Delta({\bf R,s})=\sum_{\bf k}
\Delta({\bf R,k})\textrm{e}^{\textrm{i}{\bf k\cdot s}}, 
\ \ \ \ 
\Delta({\bf R,k})=\int\textrm{d}{\bf s}\Delta({\bf R,s})
\textrm{e}^{-\textrm{i}{\bf k\cdot s}}, 
\end{equation}
with ${\bf R}=\frac{1}{2}({\bf r+r}^\prime)$. The wave functions, 
$u_i({\bf r})$ and $v_i({\bf r})$, in eq.(33) are expanded in powers of 
the relative coordinates as 
\begin{equation}
u_i({\bf R}+\frac{1}{2}{\bf s})=u_i({\bf R})+\frac{1}{2}({\bf s}\cdot
\nabla_{\bf R})u_i({\bf R})+ \frac{1}{8}({\bf s}\cdot\nabla_{\bf R})^2
u_i({\bf R})+\cdots, 
\end{equation}
\begin{equation}
v_i^\ast({\bf R}-\frac{1}{2}{\bf s})=
v_i^\ast({\bf R})-\frac{1}{2}({\bf s}\cdot\nabla_{\bf R})v_i^\ast({\bf R})
+ \frac{1}{8}({\bf s}\cdot\nabla_{\bf R})^2v_i^\ast({\bf R})+\cdots,
\end{equation}
For a $d_{xy}$-wave superconductor one can utilize the following functional 
form for the interaction, 
\begin{equation}
V({\bf s})=\frac{V_d}{k_{\rm F}^4}\sum_{\bf k}
k_x^2k_y^2{\rm e}^{{\rm i}{\bf k\cdot s}}
=\frac{V_d}{k_{\rm F}^4}\partial_x^{s2}\partial_y^{s2}\delta({\bf s}).
\end{equation}
Then, substituting eqs.(35), (36) and (37) into eqs.(33) and (34), 
we find the Fourier component of the gap function, 
\begin{equation}
\Delta({\bf R,k})
=\frac{V_d}{k_{\rm F}^4}\sum_i\int{\rm d}{\bf s}\delta({\bf s})\cdot
\partial_x^{s2}\partial_y^{s2}\bigl\{\mathcal{D}_i({\bf R,s})
\textrm{e}^{-{\rm i}{\bf k\cdot s}}\bigl\}\bigl[1-2f_{\rm F}(E_i)\bigr], 
\end{equation}
where
$$
\mathcal{D}_i({\bf R,s})=
\frac{1}{8}({\bf s}\cdot\nabla_{\bf R})^2u_i({\bf R})\cdot v_i^\ast({\bf R})
+\frac{1}{8}u_i({\bf R})\cdot({\bf s}\cdot\nabla_{\bf R})^2v_i^\ast({\bf R})
$$
\begin{equation}
-\frac{1}{4}({\bf s}\cdot\nabla_{\bf R})u_i({\bf R})\cdot
({\bf s}\cdot\nabla_{\bf R})v_i^\ast({\bf R})
\end{equation}
Assuming $\Delta({\bf R,k})\sim\Delta_0({\bf R})\hat k_x\hat k_y$ 
in the present $d$-wave case, we extract the $d_{xy}$-component from 
eq.(38) as follows, 
\begin{equation}
\Delta({\bf R,k})=-\frac{4V_d}{k_{\rm F}^2}
\sum_i\bigl[\partial_x^{s}\partial_y^{s}
\mathcal{D}_i({\bf R,s})\bigr]_{{\bf s}=0}
\bigl[1-2f_{\rm F}(E_i)\bigr]\hat k_x\hat k_y. 
\end{equation}
Thus, the gap function is obtained as 
\begin{equation}
\Delta_0({\bf R})=-\frac{4V_d}{k_{\rm F}^2}
\sum_i\bigl[\partial_x^s\partial_y^s
\mathcal{D}_i({\bf R,s})\bigr]_{{\bf s}=0}\tanh \frac{E_i}{2k_BT}, 
\end{equation}
where
$$
\bigl[\partial_x^s\partial_y^s\mathcal{D}_i({\bf R,s})\bigr]_{{\bf s}=0}
=\frac{1}{4}\Bigl\{
  \partial_x^R\partial_y^Ru_i({\bf R})\cdot v_i^\ast({\bf R})
 +u_i({\bf R})\cdot\partial_x^R\partial_y^Rv_i^\ast({\bf R})
$$
\begin{equation}
 -\Bigl[\partial_x^Ru_i({\bf R})\cdot\partial_y^Rv_i^\ast({\bf R})
 +\partial_y^Ru_i({\bf R})\cdot\partial_x^Rv_i^\ast({\bf R})\Bigr]
\Bigr\}.
\end{equation}
Note that the above gap equation is not invariant under the 
U(1)-transformation, 
\begin{equation}
\Delta_0({\bf r})\rightarrow \Delta_0({\bf r}){\rm e}^{
{\rm i}\chi({\bf r})}, \ \ 
u_\alpha({\bf r})\rightarrow u_\alpha({\bf r}){\rm e}^{
\frac{1}{2}{\rm i}\chi({\bf r})}, \ \ 
v_\alpha({\bf r})\rightarrow v_\alpha(
{\bf r}){\rm e}^{-\frac{1}{2}{\rm i}\chi({\bf r})}. 
\end{equation} 
The U(1)-symmetry may be recovered by replacing the derivatives in (A.10) 
with \lq\lq covariant" derivatives, 
$\partial_\alpha \rightarrow \partial_\alpha-\frac{{\rm i}}{2}
\phi_\alpha$ where $\phi_\alpha\equiv\partial_\alpha\phi$, $\phi$ being the 
phase of $\Delta_0({\bf r})$, i.e., $\Delta_0({\bf r})=|\Delta_0({\bf r})|
{\rm e}^{{\rm i}\phi({\bf r})}$~\cite{rf:12}. Furthermore, introducing the 
symmetrization for the differential operations,
 $\partial_x\partial_y\rightarrow \{\partial_x,\partial_y\}$, 
we rewrite the gap equation (A.9) as follows, 
$$
\Delta_0({\bf r})=-\frac{V_d}{k_{\rm F}^2}\sum_i
\Bigl[\{\partial_x-\frac{{\rm i}}{2}\phi_x,\partial_y
-\frac{{\rm i}}{2}\phi_y\}u_i({\bf r})\cdot v_i^\ast({\bf r})
+u_i({\bf r})\cdot\{\partial_x-\frac{{\rm i}}{2}\phi_x,\partial_y-
\frac{{\rm i}}{2}\phi_y\}v_i^\ast({\bf r})
$$
\begin{equation}
-(\partial_x-\frac{{\rm i}}{2}\phi_x)u_i({\bf r})\cdot (\partial_y
-\frac{{\rm i}}{2}\phi_y)v_i^\ast({\bf r})-(\partial_y
-\frac{{\rm i}}{2}\phi_y)u_i({\bf r})\cdot
(\partial_x-\frac{{\rm i}}{2}\phi_x)v_i^\ast({\bf r})\Bigr]
\tanh\frac{E_i}{2k_BT}.
\end{equation}
Here, we change the notation, ${\bf R}\rightarrow {\bf r}$. 
Suppose that the wave functions, $u_i$ and $v_i$, are expressed in 
terms of the non-trivial phase $\phi({\bf r})$ as 
\begin{equation}
\left\{
\begin{array}{l}
u_{n\mu}({\bf r})=\hat u_{n\mu}({\bf r}){\rm e}^{{\rm i}(\frac{1}{2}-\mu)
\phi({\bf r})} \\
v_{n\mu}({\bf r})=\hat v_{n\mu}({\bf r}){\rm e}^{-{\rm i}(\frac{1}{2}+\mu)
\phi({\bf r})}
\end{array}\right.,
\end{equation}
with $i=(n,\mu)$. The new wave functions, $\hat u_{n\mu}({\bf r})$ and 
$\hat v_{n\mu}({\bf r})$, are assumed to have only trivial phases. Then, 
substituting eq.(45) into eq.(46), we obtain the 
gap equation in terms of $\tilde u_{n\mu}$ and $\tilde v_{n\mu}$, 
$$
\Delta_0({\bf r})=-{\rm e}^{{\rm i}\phi({\bf r})}\cdot
\frac{V_d}{k_{\rm F}^2}\sum_{n\mu}
\Bigl[\{\partial_x-\frac{{\rm i\mu}}{2}\phi_x,\partial_y
-\frac{{\rm i\mu}}{2}\phi_y\}\tilde u_{n\mu}({\bf r})\cdot 
\tilde v_{n\mu}^\ast({\bf r})
$$
$$
+\tilde u_{n\mu}({\bf r})\cdot\{\partial_x+\frac{{\rm i\mu}}{2}\phi_x,
\partial_y+\frac{{\rm i\mu}}{2}\phi_y\}\tilde v_{n\mu}^\ast({\bf r})
-(\partial_x-\frac{{\rm i}\mu}{2}\phi_x)\tilde u_{n\mu}({\bf r})\cdot 
(\partial_y+\frac{{\rm i}\mu}{2}\phi_y)\tilde v_{n\mu}^\ast({\bf r})
$$
\begin{equation}
-(\partial_y
-\frac{{\rm i}\mu}{2}\phi_y)\tilde u_{n\mu}({\bf r})\cdot
(\partial_x+\frac{{\rm i}\mu}{2}\phi_x)\tilde v_{n\mu}^\ast({\bf r})\Bigr]
\tanh\frac{E_{n\mu}}{2k_BT}, 
\end{equation}
which indicates that the gap function $\Delta_0({\bf r})$ have the 
non-trivial phase, $\phi({\bf r})$.

\twocolumn

 \begin{figure}
\includegraphics[width=5cm,clip]{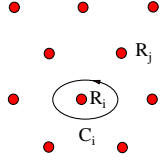}
\caption{\label{Fig.1} Closed path $C_i$ around the vortex located at 
${\bf R}_i$. Solid circles denote the vortex cores.}
 \end{figure}
%
%
%
 \begin{figure}
 \includegraphics[width=6cm,clip]{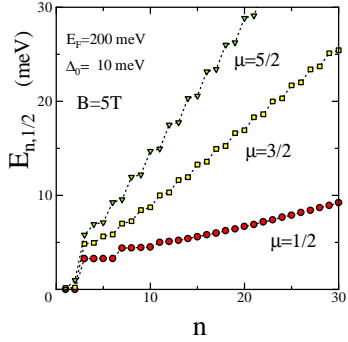}
 \caption{\label{Fig.2} Energy eigen-values for the quantum numbers 
   $\mu=\frac{1}{2}, \frac{3}{2},\frac{5}{2}$ at $B=5$T in the low energy 
   region. The eigen-value, $E_{1\mu}$, has zero-value within numerical 
   errors.}
 \end{figure}

 \begin{figure}
 \includegraphics[width=6cm,clip]{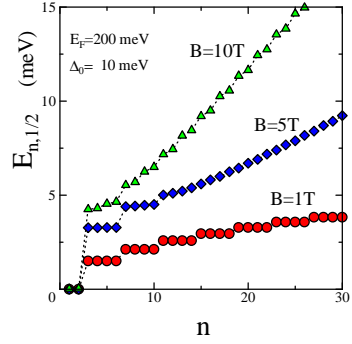}
 \caption{\label{Fig.3}Energy eigen-values of $\mu=\frac{1}{2}$-band for 
   the flux densities, $B=1, 5,10$T. The relation $\hbar\omega_c\ll 
   \Delta_0$ is satisfied in these flux densities. The zero-energy state 
   appears in this case.}
 \end{figure}
%
%
 \begin{figure}
 \includegraphics[width=6cm,clip]{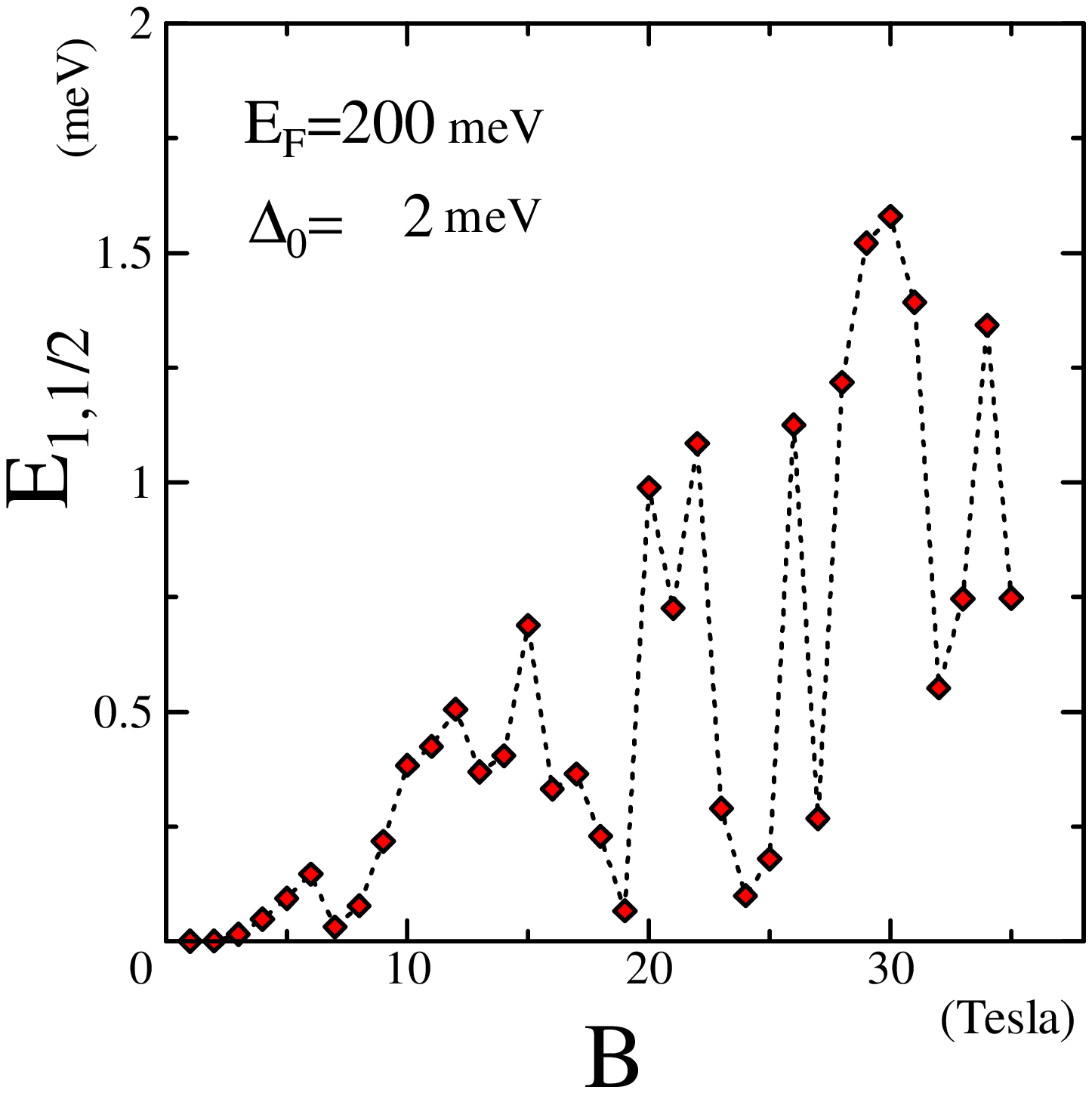}
 \caption{\label{Fig.4}Dependence of the lowest excitation energy on the 
flux density $B$. In the weak field region, where $\hbar\omega_c\ll 
\Delta_0$ is satisfied, the zero-energy quasi-particle state appears. 
However, it changes to a gapped state above some value of $B$. In the strong 
field region satisfying the relation $\hbar\omega_c > \Delta_0$ the quantum 
oscillations appear in the quasi-particle excitation energy.}
 \end{figure}



\bigskip\noindent
\begin{thebibliography}{99}
\bibitem{rf:1} G. E. Volovik: JETP Lett. \textbf{58}, (1993) 469. 
\bibitem{rf:2} Y. Wang and A. H. Macdonald:
      Phys. Rev. \textbf{B 52} (1995) R3876.
\bibitem{rf:3} S. H. Simon and P. A. Lee:
   Phys. Rev. Lett. \textbf{78} (1997) 1548.
\bibitem{rf:4} L. P. Gorkov and J. R. Schrieffer:
             Phys. Rev. Lett. \textbf{80} (1998) 3360.
\bibitem{rf:5} K. Yasui and T. Kita:
             Phys. Rev. Lett. \textbf{83} (1999) 4168.
\bibitem{rf:6} K. Krishna, N. P. Ong, Q. Li, D. Gu and N. Koshizuka:
   Science \textbf{277} (1997) 83.
\bibitem{rf:7} P. W. Anderson: cond-mat/9812063.
\bibitem{rf:8} M. Franz and Z. Te\v{s}anovi\'{c}:
    Phys. Rev. Lett. \textbf{84}, 554 (2000); \textbf{87} (2001) 257003-1.
\bibitem{rf:9} N. B. Kopnin and V. M. Vinokur:
   Phys. Rev. \textbf{B 62} (2000) 9770.
\bibitem{rf:10} J. Ye, Phys. Rev. Lett. \textbf{86} (2001) 316.
\bibitem{rf:11} A.S. Mel'nikov: 
   Phys. Rev. Lett. \textbf{86} (2001) 4108.
\bibitem{rf:12} O. Vafek, A. Melikyan, M. Franz and Z. Te\v{s}anovi\'{c}:
    Phys. Rev. \textbf{B 63} (2001) 134509.
\bibitem{rf:13} We assume that the BdG equation (1) is invariant under 
a trivial phase transformation, $\phi({\bf r})\rightarrow \phi({\bf r})+
\chi({\bf r})$, $u_\alpha({\bf r})\rightarrow u_\alpha({\bf r}){\rm e}^{
\frac{1}{2}{\rm i}\chi({\bf r})}$, $v_\alpha({\bf r})\rightarrow v_\alpha(
{\bf r}){\rm e}^{-\frac{1}{2}{\rm i}\chi({\bf r})}$, with $\chi({\bf r})$ 
being a single-valued function. 
%
\bibitem{rf:14} In order to merely eliminate the phase factor in the BdG 
equation arbitrary functions are allowed for $\phi_e({\bf r})$ and $\phi_h
({\bf r})$ as far as the condition $\phi_e+\phi_h=\phi$ is fulfilled. In the 
Franz-Te\v{s}anovi\'{c} transformation the special phases are chosen for 
$\phi_e$ and $\phi_h$ as $\phi_e({\bf r})=\phi_A({\bf r})$ and $\phi_h
({\bf r})=\phi_B({\bf r})$, where $\phi_A({\bf r})$ ($\phi_B({\bf r})$) is 
the non-trivial phase coming from the vortices on the 
A-sublattice (B-sublattice)~\cite{rf:8}. The wave functions derived in 
ref.[8], using the Franz-Te\v{s}anovi\'{c} transformation, does not show the 
correct AB effect in the $N$-vortex state. The detailed comment for the 
Franz-Te\v{s}anovi\'{c} transformation will be given in a separate paper.
%
\bibitem{rf:15} The operator $\mathcal{P}(C)$ which moves a function 
from ${\bf r}_1$ to ${\bf r}_2$ along a path $C$ is expressed as 
$\mathcal{P}(C)={\rm exp}\int_0^1{\rm d}\tau\frac{{\rm d}{\bf s}}{d\tau}
\cdot\nabla_{\bf r}$, where the equation of path $C$ is given as 
$C=\bigl\{{\bf s}(\tau)|0\leq \tau \leq 1, {\bf s}(0)={\bf r}_1, {\bf s}(1)
={\bf r}_2\bigr\}$.
%
\bibitem{rf:16} F. Gygi and M. Schl\"uter:
     Phys. Rev. \textbf{B 43} (1991) 7609.
\bibitem{rf:17} This point was overlooked in previous works.
In ref.[8] the terms containing $\partial_i\partial_j\phi$ are neglected 
on the assumption that these terms are higher-order derivatives and then 
they are small. In such an approximation the problem about the order of 
differentiations does not arise. However, the assumption is incorrect. Since 
$\partial_i\partial_j\phi\propto B$, as shown in our calculations, these terms 
should be considered as the 0th order ones in the strong field region. 
%
\bibitem{rf:18} The topological singularity given in eq.(5) is expanded 
into the Fourier series as 
$$
\nabla\times\nabla\phi({\bf r})|_z=2\pi\sum_i^N\delta({\bf r-R}_i)
=2\pi m\sum_{\bf G}{\rm e}^{{\rm i}{\bf G\cdot r}}
=2\pi m + 2\pi m\sum_{{\bf G}\not=0}{\rm e}^{{\rm i}{\bf G\cdot r}},
$$
for a regular flux-line-lattice state, where ${\bf G}$ denotes the reciprocal 
lattice vector. In the approximation in which the terms of ${\bf G}\not=0$ 
are neglected one gets the approximate result given in eq.(24). 
%
\bibitem{rf:19} T. Koyama and M. Tachiki:
            Phys. Rev. \textbf{B 53} (1996) 2662.
\end{thebibliography}
\end{document}